\documentclass[preprint,12pt]{elsarticle}
\usepackage{amssymb}
\usepackage{amsmath}

\journal{Photoacoustics}

\begin{document}

\begin{frontmatter}

\title{An integrated micro-thermometer for mid-infrared photothermal spectroscopy of dense materials built in a CMOS $200~\mathrm{mm}$ pilot line\tnoteref{t1}}
\tnotetext[t1]{This work is part of the IPCEI Microelectronics and Connectivity and was supported by the French Public Authorities within the frame of France 2030. It was also supported by the CARNOT Institute LETI.}

\author[1]{Sandy Mathew}

\author[1]{Sonia Messaoudène}

\author[1]{Adrien Poizat}

\author[1]{\'Eléa Bourliaud}

\author[1]{Marion Volpert}

\author[1]{Jules Skubich}

\author[1]{Stanislas Lhomme}

\author[1]{Bertrand Bourlon}

\author[1]{Kevin Jourde\corref{cor1}}
\ead{kevin.jourde@cea.fr}
\cortext[cor1]{Corresponding author}

\affiliation[1]{
organization={Univ. Grenoble Alpes, CEA, LETI},
addressline={F38054}, 
city={Grenoble},
country={France}
}

\begin{abstract}
Indirect photothermal infrared spectroscopy of dense materials typically relies on photoacoustic cells for detecting acoustic waves generated by light-induced heating. This detection approach can be challenging to implement as it is highly sensitive to the sealing quality of the interface between the cavity and the sample, as well as to the properties of the gas within the cavity. Here, we present a simplified and more robust method. This approach relies on direct photothermal temperature measurement using an optimized platinum microsensor fabricated on a thin silicon substrate that acts as a thermally transparent interface with the sample. The sensor design, fabrication in a $200~\mathrm{mm}$ CMOS pilot line, assembly on a readout PCB, and experimental characterization are reported. The component achieves a signal-to-noise ratio close to $1000$ making it well-suited for miniaturized and embedded infrared spectroscopic applications.
\end{abstract}

\begin{graphicalabstract}
\includegraphics[width=1.0\columnwidth]{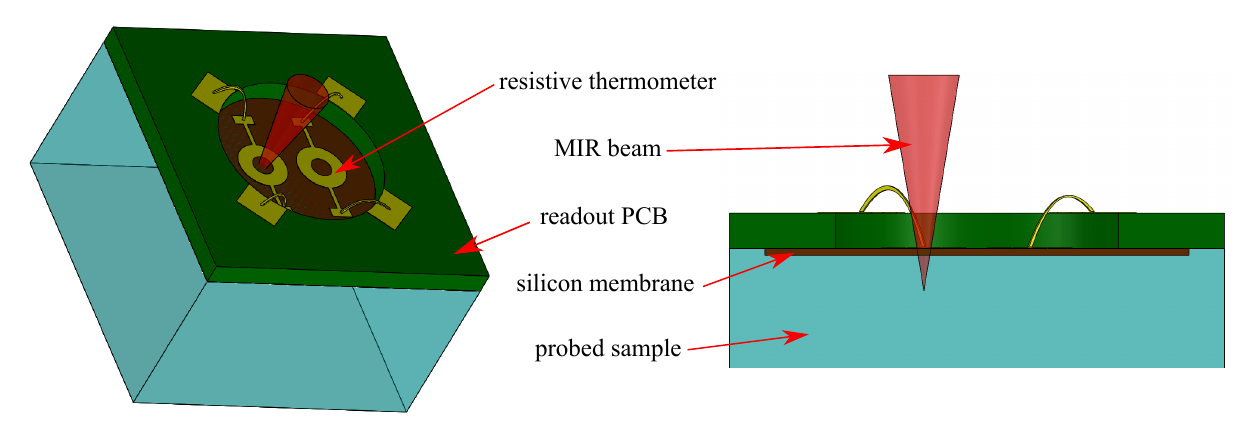}
\end{graphicalabstract}

\begin{highlights}
\item Photothermal infrared spectroscopy of liquids with a temperature microsensor.
\item Platinum micro-thermometers integrated with thin silicon wafer in a CMOS line.
\item Direct photothermal measurement of diluted glucose with high signal-to-noise ratio.  
\end{highlights}


\end{frontmatter}

\section{Introduction}
\label{sec:introduction}

Infrared (IR) spectroscopy is a widely used method for material characterization. It relies on measuring the spectral fingerprint associated with the absorption of infrared radiation by molecular rovibrational energy transitions \cite{james_a_de_haseth_introduction_2007}. The mid-infrared (MIR) spectral range is of particular interest due to its specific absorption by molecules in gas, liquid, or solid state \cite{krebbers_mid-infrared_2022, baker_using_2014}. It therefore lends itself to diverse applications to identify and quantify a wide variety of chemical species such as for environmental and industrial gas sensing \cite{krzempek_review_2019, lim_ultrasensitive_2024}, microelectronics characterization \cite{chabal_applications_2002}, disease diagnosis and monitoring \cite{de_bruyne_applications_2018}, forensic \cite{flores_development_2025}, industrial control and monitoring \cite{kau-wacht_laser-based_2025}. However, IR spectroscopy is mostly confined to laboratory-level analysis as it relies on expensive and bulky instruments. Recent advances in quantum cascade lasers (QCL) \cite{scalari_30_2024, lepage_hybrid_2025} have partially addressed this limitation by setting the way to future compact, efficient, and low-cost IR sources. In this work, we present a novel technique based on direct photothermal MIR spectroscopy that can be employed on liquid, soft-matter or solid samples. Our approach enables a more compact and cost-effective implementation, improving the accessibility of MIR spectroscopy.

Photothermal infrared (PTIR) spectroscopy involves recovering an absorption spectrum by measuring temperature associated to the heat released when a sample is exposed to a modulated IR beam \cite{rosencwaig_theory_1976, mcdonald_generalized_1978, hu_generalized_1999}. It offers major advantages such as the ability to characterize optically opaque samples, insensitivity to sample transmission and scattering features, along with a simplified detection scheme. However, it is sensitive to sample thermomechanical parameters which in some cases may render the data interpretation difficult. Several experimental approaches based on photothermal spectroscopy have been developped \cite{tam_applications_1986, proskurnin_11_2014}. Among the most widely used are direct photoacoustic spectroscopy (detection of acoustic waves generated within sample due to thermal expansion or compression) \cite{fathy_direct_2022}, indirect photoacoustic spectroscopy (acoustic waves arising from sample surface temperature variations within a cavity are detected) \cite{perondi_minimalvolume_1987, falkhofen_quartz_2024}, deflection spectroscopy (local refractive index fluctuations due to sample surface temperature variations are detected) \cite{vlk_spatial_2024}, and photothermal radiometry (radiative thermal emission is measured) \cite{fuente_simultaneous_2011}. Interest in PTIR-based techniques has grown over the past decade because they are well suited for the development of embedded IR spectroscopy systems, and could provide a solution for non-invasive monitoring of humans physiological parameters \cite{kottmann_mid-infrared_2016, coutard_neogly_2024, uluc_non-invasive_2024}.

While most of these methods rely on indirect detection of light-induced heating, an alternative approach is the direct measurement of the temperature variations using resistive thermometers. This concept was demonstrated in \cite{hammiche_photothermal_1999} where a resistive probe was integrated close to polymeric samples in a Fourier transform infrared spectroscopy (FT-IR) setup. In a later development \cite{katzenmeyer_mid-infrared_2015}, the contact tip of an atomic force microscope was replaced with a resistive thermal probe to directly measure photothermal response. However, such platforms are sophisticated, expensive, and difficult to scale. Here, we demonstrate a different approach based on a miniaturized integrated micro-thermometer to conduct photothermal measurements. The sensor could be mounted, for example, on a microfluidic cartridge (similar to \cite{davaji_microscale_2019}) or brought into direct contact with a sample, as in a smartwatch. This device was fabricated using standard processes in a $200~\mathrm{mm}$ CMOS pilot line. Compared to other photothermal techniques, the attractiveness of our approach lies in its simplicity in design and operation as well as fabrication compatibility with standard semiconductor device manufacturing processes. Furthermore, combined with a QCL (or a QCL bundle) as the optical source, it paves the way for compact embedded IR spectroscopy systems.

\begin{figure}[h!]
\centering
\includegraphics[width=1.0\columnwidth]{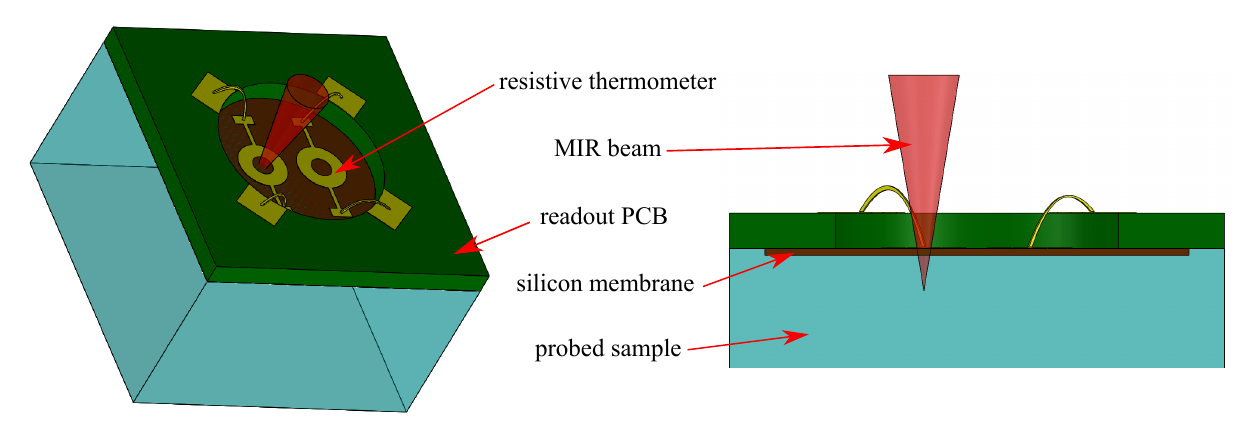}
\caption{Simplified views of the THERMOMIR operating principle. The 3D illustration represents the micro-thermometer on a thin silicon substrate (in red) integrated to a readout PCB (in dark green) probing a liquid sample underneath (in blue). The right illustration is a section view of the 3D model.}
\label{fig:component_principle}
\end{figure}

The approach described in this manuscript is the following (see Fig.~\ref{fig:component_principle})~: a thin substrate (between $25~\mathrm{\mu m}$ and $100~\mathrm{\mu m}$), made of an IR transparent material, is placed in contact with a sample. The sample can be a flowing liquid, a solid, or a soft material. A planar resistive thermometer is fabricated on the opposite face of the substrate that is in contact with the ambient air. The sample is exposed to an IR beam passing through the substrate. The heat released by the sample diffuses toward the front side of the substrate containing the resistive thermometer, inducing a change in its resistance. The thermometer geometry is optimized to efficiently detect the photothermal response. In the following sections, the component design, fabrication, assembly, and characterization are successively described. The component is hereafter referred to as THERMOMIR. 

\section{Modelling and design}
\label{sec:modelling_and_design}

\begin{figure}[h!]
\centering
\includegraphics[width=1.0\columnwidth]{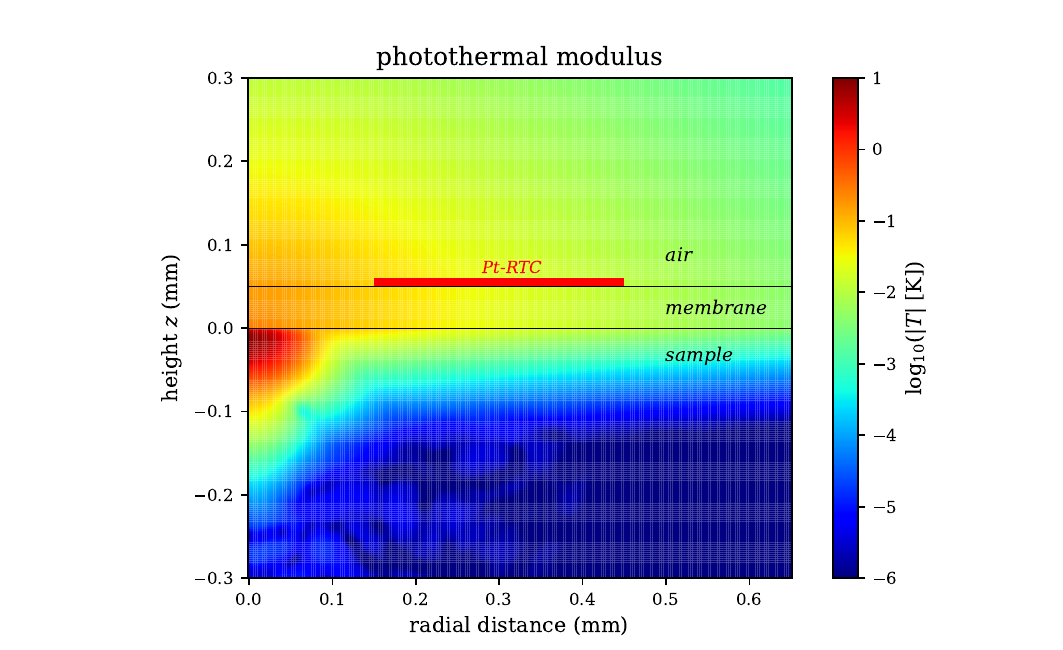}
\caption{2D section view of the simulated photothermal temperature modulus profile. Only half of the sensor is presented (because the sensor has a revolution symetry around the $z$ axis). The component is represented as a thick red line. It corresponds to the geometry retained at the end of Sec.~\ref{sec:modelling_and_design}, and testing conditions similar as those chosen in Sec.~\ref{sec:results_and_discussion}. The $\mathrm{Si}$ membrane is $50~\mathrm{\mu m}$ thick and the sample is water. The heating source is modeled as a Gaussian beam of power $9~\mathrm{mW}$ with intensity modulated at $237~\mathrm{Hz}$ and of beam radius $75~\mathrm{\mu m}$. The heating source spatial profile is deduced from the Beer-Lambert law for a monochromatic source set at $9.60~\mathrm{\mu m}$.}
\label{fig:photothermal_temperature_profile}
\end{figure}

The design criteria for THERMOMIR were established based on simulation results. The component was modeled using the finite element method\footnote{The simulations were performed with COMSOL Multiphysics 5.4-6.3 and the PARADISIO solver.} to estimate the amplitude and the spatial distribution of the thermal diffusion at substrate surface as well as the thermometer electrical resistive response under modulated optical excitation. A three-layer system consisting of water (the sample), solid substrate, and air was simulated. The source was modeled as a Gaussian beam with a wavelength of $9.60~\mathrm{\mu m}$ with an exponentially decreasing intensity with depth due to absorption in water according to the Beer-Lambert law. The thermal and electrical phenomena were modeled by solving the heat diffusion and electrostatic equations. Fig.~\ref{fig:photothermal_temperature_profile} shows a cross-sectional view of the scalar temperature field in the simulated configuration. For standard operating conditions (detailed in Fig.~\ref{fig:photothermal_temperature_profile}), the signal amplitude is expected to be approximately $100~\mathrm{mK}$. The device becomes relevant for spectroscopy applications if relative temperature variations ranging from one-thousandth to one-hundredth of the signal total amplitude can be detected.

In the following subsections, we detail our choices for the substrate and the thermometer designs.

\subsection{Thin silicon substrate}
\label{subsec:membrane}

The sensor spatial dimensions are closely related to the optical intensity modulation frequency f [$\mathrm{Hz}$]. Modulating $f$ is of prime interest when performing photothermal spectroscopy measurements as it allows for depth profiling of the sample \cite{rosencwaig_theory_1976}. The temperature detected at the sample surface results from heat diffusing from a given depth within the sample to the surface. This depth corresponds to the sample thermal diffusion length $\mu ~\mathrm{[m]}$~:

\begin{equation}
\mu = \sqrt{\dfrac{k}{\pi f\rho C_p}} \mathrm{~~~,}
\label{eq:diffusion_length}
\end{equation}
where $k~\mathrm{[W.m^{-1}.K^{-1}]}$ is the thermal conductivity,  $\rho~\mathrm{[kg.m^{-3}]}$ is the density,  and $C_p~\mathrm{[J.kg^{-1}.K^{-1}]}$ is the specific heat capacity of the material. It is assumed that the sample is thermally thick, meaning that the sample thickness $L~\mathrm{[m]}$ and the modulation frequency $f$ are chosen such that $L \gg \mu$. The second critical length scale is the optical penetration depth $\delta ~[m] = 1 / \beta$, where $\beta~\mathrm{[m^{-1}]}$ is the so-called optical absorption coefficient (or absorbance). It defines the spatial profile of the heat source generated by optical excitation, following the Beer-Lambert law~:

\begin{equation}
S(r, z, t) = \beta I(r, t) \mathrm{e}^{+\beta z} \mathrm{~~~,}
\label{eq:heat_source}
\end{equation}
where $S~\mathrm{[W.m^{-3}]}$ is the heat source induced by an electromagnetic wave, with an intensity $I~\mathrm{[W.m^{-2}]}$ at the sample surface ($z = 0$). Two regimes can be distinguished based on the critical frequency $f_t$, at which the optical penetration depth and the thermal diffusion length become equal i.e $\delta = \mu_{sample}$. Assuming an optically opaque sample ($L \gg \delta$), when $\mu_{sample} >\delta$, the heat detected at the surface is mainly sensitive to the sample thermomechanical properties. Conversely, when $\mu_{sample} < \delta$, the measurement depends both on the sample thermomechanical properties and its optical absorption. In this regime, the sample absorbance can be probed at different depths by selecting appropriate modulation frequencies. A complete characterization therefore requires measurements over a frequency range encompassing $f_t$. For a liquid (mostly composed of water), at a wavelength of $9.60~\mathrm{\mu m}$, $\delta \approx 20 ~\mathrm{\mu m}$, and $f_t \approx 200 ~\mathrm{Hz}$. Consequently, data are typically acquired for frequencies ranging from $10~\mathrm{Hz}$ to $1~\mathrm{kHz}$. 

The frequency range being set, it is possible to define the substrate thickness $l~\mathrm{[m]}$ and material based on its thermomechanical properties. To ensure efficient heat diffusion from the sample to the substrate surface containing the microsensor, the condition $l \ll \mu_{susbtrate}$ must be satisfied. The substrate should be thin, to have a high thermal conductivity, and to have a low thermal capacity and density. It should also be optically transparent to maximize IR beam transmission toward the sample, and avoid the generation of non-sample-specific signals. Silicon was therefore selected as the substrate material. It fulfills all these requirements and is compatible with standard microfabrication processes. More precisely, at $9.60~\mathrm{\mu m}$, $\delta_{Si} \approx 25 ~\mathrm{mm}$, and $\mu_{Si} \approx 5 / \sqrt{f \mathrm{[Hz]}}~\mathrm{mm}$ \cite{shanks_thermal_1963, james_experimental_1982}. For the upper limit of the modulation frequency range $f = 1~\mathrm{kHz}$, $\mu_{Si} \approx 160 ~\mathrm{\mu m}$. Therefore, a substrate thickness $l$ below $100~\mathrm{\mu m}$ is desirable to ensure efficient heat diffusion within the operating frequency range. However, at such thicknesses, Si substrates become difficult to handle, which constrains the fabrication and integration steps (discussed in Sec.~\ref{sec:fabrication_and_integration}). 

Si substrates of three different thicknesses, $l \in [25, 50, 100] ~\mathrm{\mu m}$, were fabricated using the process described in Sec.~\ref{sec:fabrication_and_integration}. While the $25~\mathrm{\mu m}$ substrate was prone to breakage, the two other thicknesses could be successfully integrated and tested. For the component tested in Sec.~\ref{sec:results_and_discussion}, $l = 100~\mathrm{\mu m}$.

\subsection{Platinum micro-thermometer}
\label{subsec:thermistance}

A resistance temperature detector (RTD) based on thin-film metal was selected as the sensing element. Its operating principle relies on the linear variation of electrical resistance with temperature. RTD devices are characterized by the material temperature coefficient of resistance (TCR) $\alpha ~\mathrm{[K^{-1}]}$, which is included in the first-order approximation of the temperature-dependent resistance~:

\begin{equation}
R(T) = R_0 (1 + \alpha (T - T_0)) \mathrm{~~~,}
\label{eq:rtd_response}
\end{equation}
where $R~\mathrm{[\Omega]}$ is the RTD resistance, $T~\mathrm{[K]}$ is the temperature, and $R_0 = R(T_0)$. $T_0 = 298.15~\mathrm{K}$ is the reference temperature. A platinum (Pt) RTD was prefered for its high $\alpha$ value, linear response, stability, and ease of fabrication and characterization \cite{sousa_development_2021}.

\begin{figure}[h!]
\centering
\includegraphics[width=1.0\columnwidth]{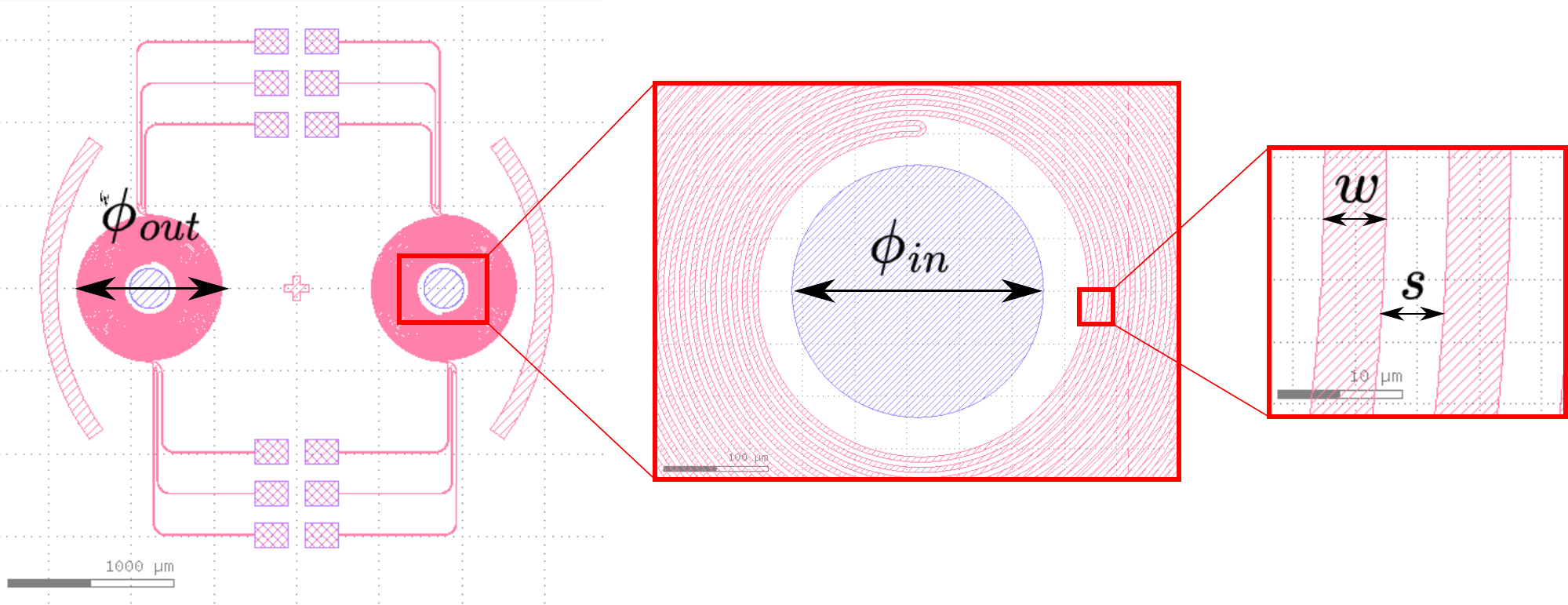}
\caption{Schematic of the micro-sensor design. Each unit consists of two identical spirally-wound microsensors, only one of which is exposed to the Gaussian beam through the central region (blue). This configuration enables differential measurements, whereby non-specific signals are directly eliminated, allowing measurement of only the photothermal temperature response.}
\label{fig:thermo_geom}
\end{figure}

A spirally-wound trace geometry was chosen for our RTD device (see Fig.~\ref{fig:thermo_geom}). The shape was designed by taking into account the isotropic behaviour of heat diffusion resulting from the optical excitation configuration. The traces are arranged in a spiral around a central region of a given diameter, allowing the Gaussian MIR beam to pass through. Indeed, it is desirable to minimize direct interaction between the beam and the RTD while maximizing heat diffusion from the sample toward the RTD.

The parameters affecting the thermometer performance are the inner area diameter $\phi_{in} ~\mathrm{[m]}$, the RTD trace width $w~\mathrm{[m]}$ and thickness $e~\mathrm{[m]}$, the spacing between traces $s~\mathrm{[m]}$, and the resistance $R_0$. The outer diameter $\phi_{out} ~\mathrm{[m]}$ is then determined from the resulting geometry. The choice of inner diameter $\phi_{in}$ is a compromise between several constraints. Ideally, it should be smaller than the thermal diffusion length in silicon to maximize the overlap between the thermometer traces and the spatial thermal profile. It is also desirable to concentrate the beam over a small area to locally induce large temperature changes. However, sample overheating must be avoided, and beam alignment should remain practical. In this case, $\phi_{in}$ was set to $240 ~\mathrm{\mu m}$. For the trace width $w$, spacing between traces $s$, and the thickness $e$, values commonly reported in the literature were selected \cite{sousa_development_2021, song_integrated_2020}. The values of $w$ and $e$ were set to either $5$ or $10~\mathrm{\mu m}$, while $e$ was fixed at $200~\mathrm{nm}$. For a given nominal resistance $R_0$, as mentioned previously, it is advantageous to concentrate the spiral traces close to the beam zone (see figure Fig.~\ref{fig:photothermal_temperature_profile}), thereby minimizing $w$, $s$, and $e$. The value of $R_0$ was set by adjusting the trace length (or outer diameter) to match the electronic readout specifications. Three values were therefore selected for $R_0$~: $2$, $8$, and $24~\mathrm{k \Omega}$.

As shown in Fig.~\ref{fig:thermo_geom}, a differential configuration was chosen. It consists of two identical and interchangable components placed side by side. One serves as the active component, surrounding the beam passing through the central region, while the other acts as a reference. The measurement consists of subtracting the signal from the two RTDs, enabling an improved sensitivity.

Nine RTD variants were fabricated, as shown in Fig.~\ref{fig:wafer_and_component_pictures}. For the sake of brevity, only measurements performed on one variant (presented in Fig.~\ref{fig:thermo_geom}) are reported here. This variant features $R_0 = 8~\mathrm{k \Omega}$, $\phi_{out} = 875~\mathrm{\mu m}$, $\phi_{in} = 240~\mathrm{\mu m}$, $w = 5~\mathrm{\mu m}$, $s = 5 ~\mathrm{\mu m}$, and $e = 200 ~\mathrm{nm}$.

\section{Fabrication and Integration}
\label{sec:fabrication_and_integration}

\subsection{Cleanroom fabrication}
\label{subsec:clean_room_fabrication}

The RTD components were patterned and fabricated following standard photolithography process. However, the main challenge was the handling of a fragile substrate of thicknesses ranging from $25$ to $100~\mathrm{\mu m}$ within a conventional $200~\mathrm{mm}$ CMOS line. The issue was solved by using a succession of holders bonded to the thin substrate with thermal release films (NITTO REVALPHA$^\mathrm{TM}$). Fig.~\ref{fig:fabrication_and_assembly_process} summarizes the main cleamroom fabrication steps (steps 1 to 3)~:

\begin{figure}[h!]
\centering
\includegraphics[width=1.0\linewidth]{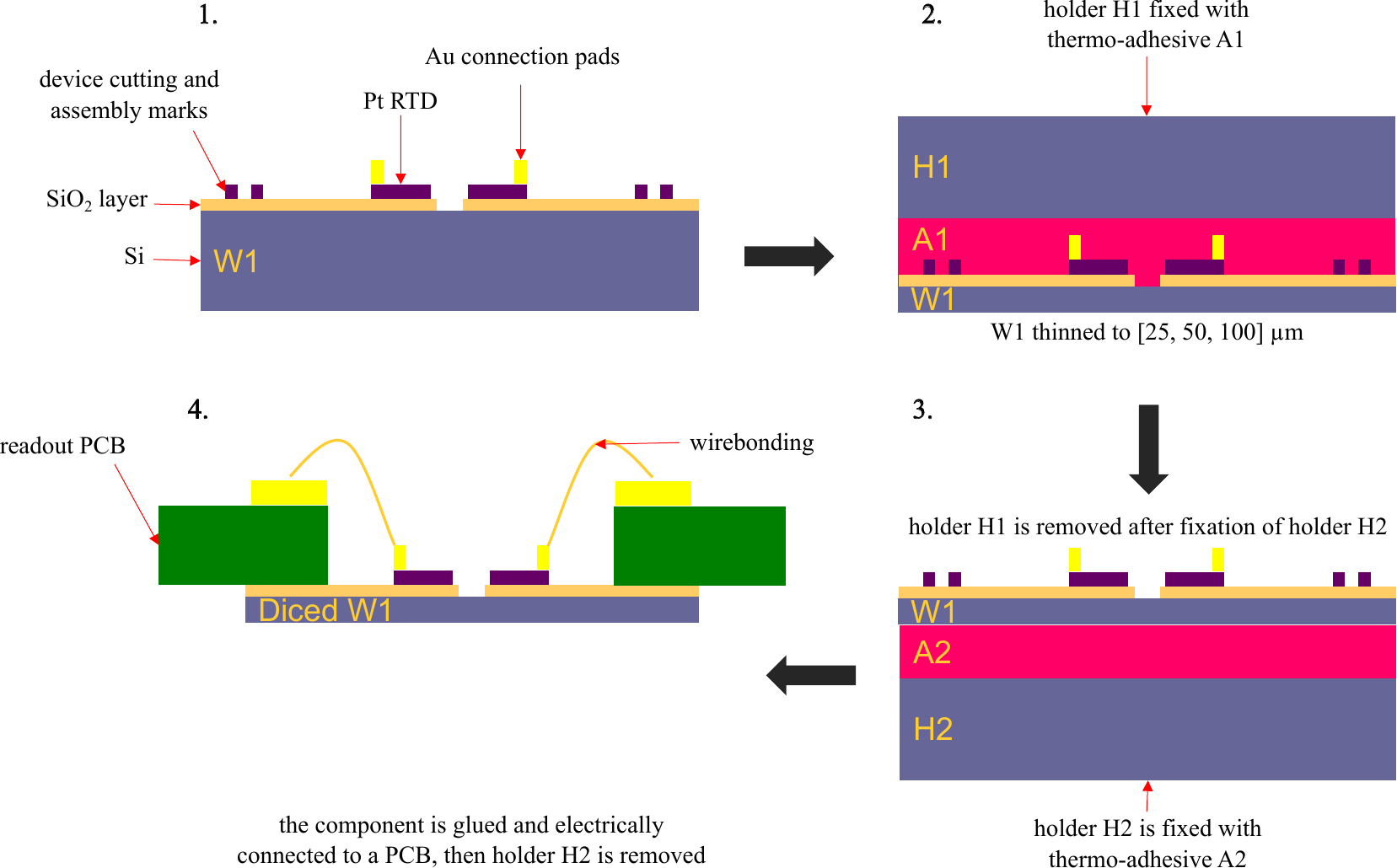}
\caption{Schematic THERMOMIR sensor fabrication process in a $200~\mathrm{mm}$ CMOS line (steps 1-3) and its integration (step 4). Step 1 consists of photolithographic patterning of the RTD device, including the $\mathrm{Ti/Pt}$ thermometer, and the $\mathrm{Au}$ pads for electrical contacts. Steps 2 and 3 involve a series of bonding-debonding processes using adhesive films and wafer holders (H1-H2). In step 4, after dicing, the component is assembled onto a PCB through gluing, wirebonding, and removal of holder H2.}
\label{fig:fabrication_and_assembly_process}
\end{figure}

\begin{itemize}
\item[Step 1~:] A thin silicon dioxide layer (typically $100~\mathrm{nm}$) is oxidized on an undoped silicon wafer (W1) to prevent current leakages. Each metal layer is then defined through successive steps of metal deposition, photolithography, and etching. The RTD thermometer consists of a $10~\mathrm{nm}$ titanium adhesion layer followed by $200~\mathrm{nm}$ of platinum layer forming the thermometer traces. Locally, a $200~\mathrm{nm}$ gold layer is also deposited on the electrical pads for wirebonding to the readout PCB. Finally, the oxide layer is removed from the thermometer center to enable IR beam transmission. 
\item[Step 2~:] A holder (H1) (a standard silicon wafer) is bonded to the front surface of W1 using a thermal adhesive (A1). The holder enables thinning W1 to its final thickness ($25-100~\mathrm{\mu m}$). The thickness is first reduced by rough mechanical grinding followed by dry etching.
\item[Step 3~:] Similar to step 2, a second holder (H2) is attached to the rear side of the wafer (W1) using another thermal adhesive (A2). Holder H1 is then removed together with A1. A1 and A2 are characterized by a threshold temperature above which the adhesion is lost. By choosing a higher threshold temperature for A2, H1 can be removed while keeping H2. In our case $T_{A1} = 170~\mathrm{^{\circ}C}$ and $T_{A2} = 200~\mathrm{^{\circ}C}$. \
\end{itemize}

The process flow described up until step 3 is compatible with wafer-level characterization (see figure Fig.~\ref{fig:wafer_and_component_pictures}). Specifically, the presence of the holder H2 enables easy handling of the stack without mechanical breakage. Moreover, the thermal sheet A2 can serve as an IR absorbing medium for wafer-level photothermal characterization.

\subsection{Device integration}
\label{subsec:device_integration}

\begin{figure}[h!]
\centering
\includegraphics[width=1.0\columnwidth]{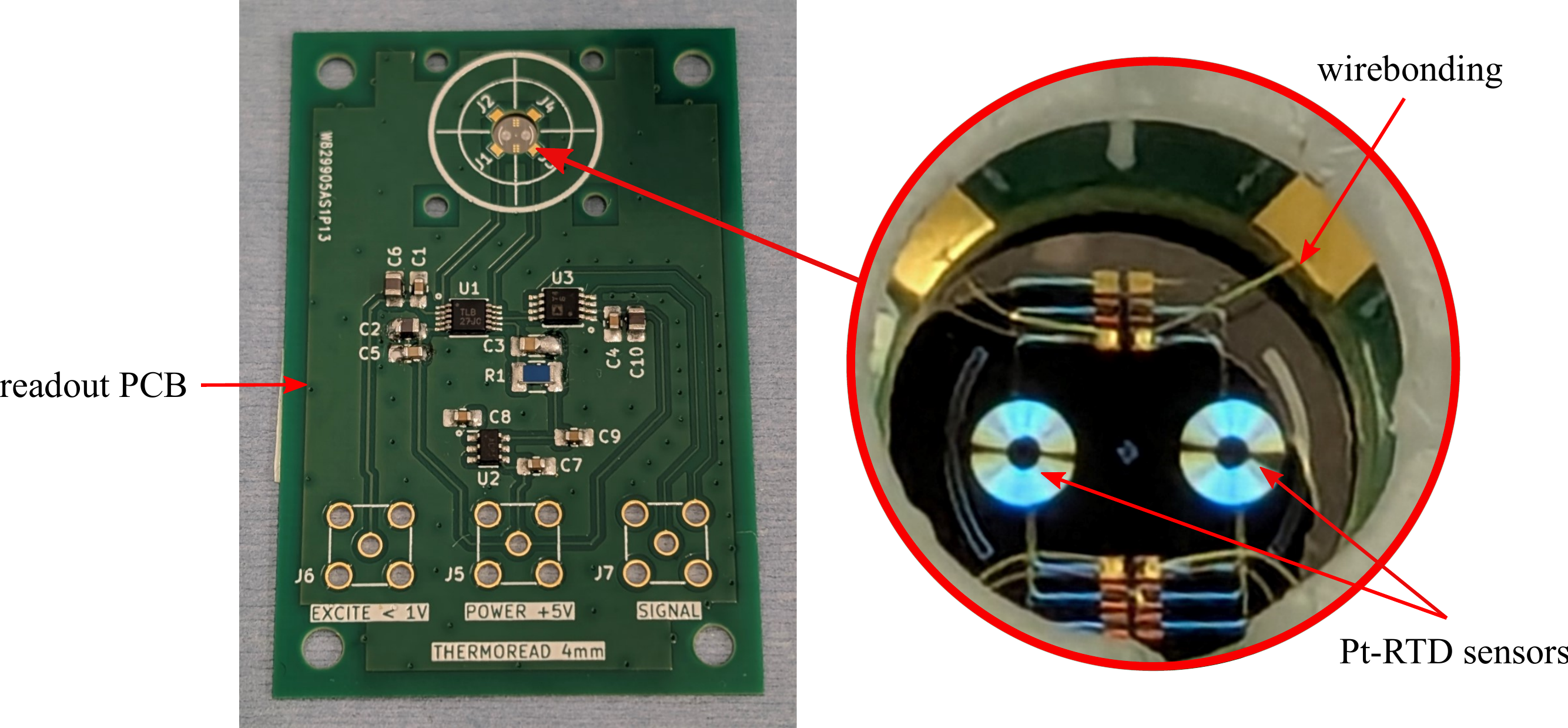}
\caption{THERMOMIR sensor with its readout PCB is shown on the left, and a close-up picture of the integrated RTD device is shown on the right.}
\label{fig:integrated_device_pictures}
\end{figure}

In order to characterize the RTD sensor at the die level, the wafer stack (Step 3 mentioned previously) was diced and assembled onto a readout PCB (see Fig.~\ref{fig:fabrication_and_assembly_process} (step 4) and Fig.~\ref{fig:integrated_device_pictures}). After dicing, die integration onto the PCB involves three additional steps~: bonding the die to the PCB, electrically interconnecting it to the PCB via wirebondings, and finally releasing the thermal adhesive and holder (A2-H2 in Fig.~\ref{fig:fabrication_and_assembly_process}). Here again, holder H2 was critical for mounting the die without the risk of mechanical breakage.

\begin{figure}[h!]
\centering
\includegraphics[width=1.0\columnwidth]{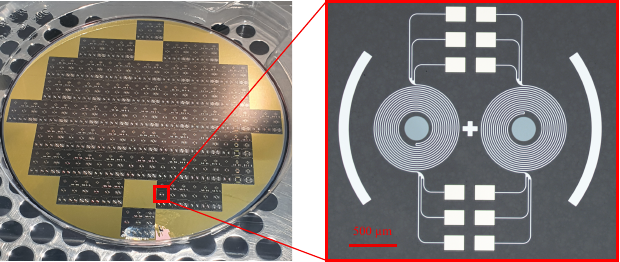}
\caption{Picture of the THERMOMIR wafer on the left, and optical microscope image of an individual sensor on the right. The cross at the center serves as an alignment mark for PCB integration.}
\label{fig:wafer_and_component_pictures}
\end{figure} 

As shown in Fig~\ref{fig:wafer_and_component_pictures}, each component has six electrical connection pads. Four of them are used for wafer-level electrical characterization, while the remaining two are reserved for the final PCB integration.

The main component of the PCB is a precision differential thermistor signal amplifier (TEXAS INSTRUMENTS INA330), used in a configuration similar to \cite{song_integrated_2020}. For a given reference voltage, it operates by measuring the difference in current flowing through the reference and measurement sensors. The resulting signal is then amplified and filtered according to the application specifications. The INA330 is optimized for resistance values around $10~\mathrm{k\Omega}$ which corresponds to the retained value of $R_0$ mentioned previously.

\section{Characterization results and Discussion}
\label{sec:results_and_discussion}

In this section, the characterization results obtained from one of the THERMOMIR sensor variants (visible on Fig.~\ref{fig:thermo_geom}), both at the wafer and device levels, are presented. The specific characteristics of this variant are listed at the end of Sec.~\ref{subsec:thermistance}. Pictures of the characterization setups are provided in \ref{ann:characterization_benches}

\subsection{Wafer-level characterization}
\label{subsec:wafer_level_charac}

The components were tested at the wafer-level using an automatic prober equipped with a temperature-controlled chuck (FORMFACTOR CM300). 

\begin{figure}[h!]
\centering
\includegraphics[width=1.0\columnwidth]{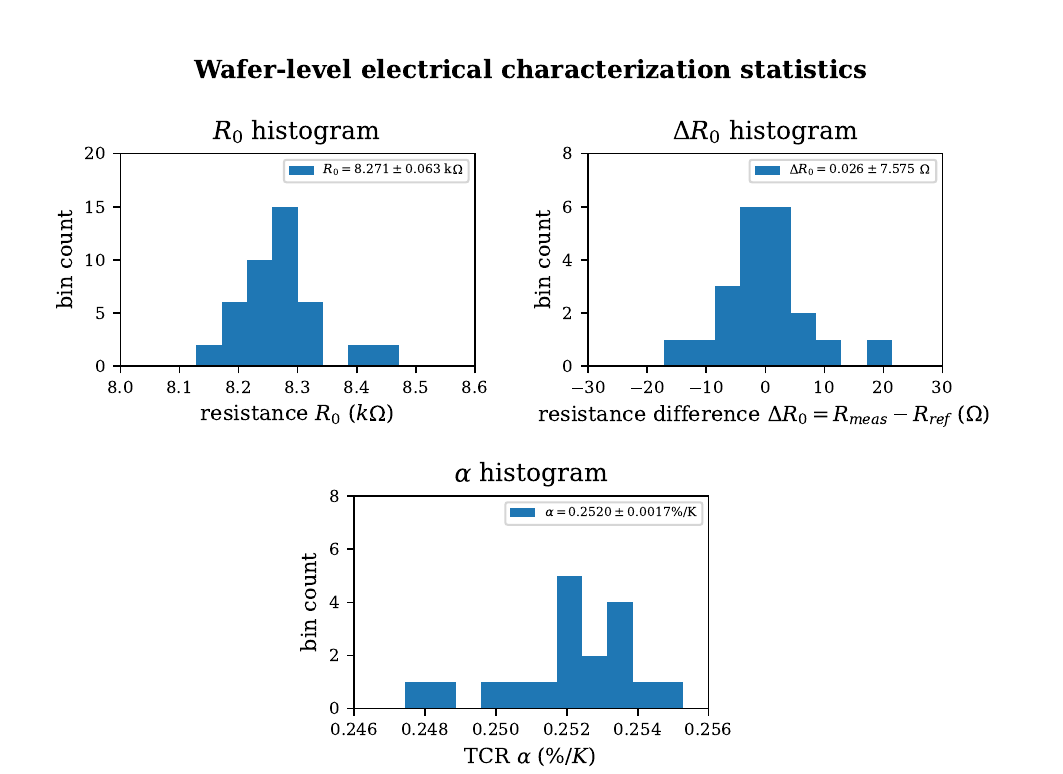}
\caption{Histograms of measured $R_0$ (\textit{top left}), $\Delta R_0 = R_{meas} - R_{ref}$ (\textit{top right}), and $\alpha$ (\textit{bottom}) obtained from wafer-scale characterization of the THERMOMIR sensors (measurement and reference). The measured $R_0$ agrees with simulation results, and $\alpha$ is compatible with values reported in the literature. The variability of $\Delta R_0$ is smaller than $R_0$ because $R_0$ histogram is affected by a wafer-scale spatial construction drift.}
\label{fig:resistance_and_tcr_histograms}
\end{figure}

First, the components were electrically characterized. Both the measurement and reference sensors were individually tested. The characterization consisted of resistance measurement using 4-point probes ($20~\mathrm{\mu m}$ tip radius beryllium-copper probes) at a fixed reference temperature ($T_0 = 298.15~\mathrm{K}$), followed by the determination of $\alpha$ (TCR) by measuring the resistance at various regulated temperatures and fitting the results using Eq.~\ref{eq:rtd_response}. A statistical analysis of these measurements is presented in Fig.~\ref{fig:resistance_and_tcr_histograms}. The measured resistance ($\sim 8.3~\mathrm{k \Omega}$) is in agreement with the numerically estimated value ($8.0 ~\mathrm{k \Omega}$). The discrepancy (under $4\%$) can be explained by the fabrication variations and difference between the actual platinum resistivity with respect to the values used in our simulations. A standard deviation of $\sim 60~\mathrm{\Omega}$ was obtained for $R_0$. The variability is smaller when comparing two adjacent sensors ($\Delta R_0 = R_{meas} - R_{ref}$)~: $\sim 8~\mathrm{\Omega}$. The difference can be explained by a wafer-scale fabrication process drift. TCR values were estimated at $\sim 0.25~\mathrm{\%.K^{-1}}$ which is in good agreement with those reported in the literature \cite{dan_effects_2023}.

Second, the component's photothermal response was measured by connecting them to an external readout PCB. A QCL was used as the excitation source, with its beam redirected to the measurement RTD central area. In this configuration, the thermal adhesive film A2 was used as the sample (shown in Fig.~\ref{fig:fabrication_and_assembly_process}). A photothermal signal was successfully measured, but showed a large variability in the photothermal signal amplitude. We hypothesize that this behaviour is due to non-uniformity or heterogeneity in the bonding of the thermal film to the wafer. While these measurements allowed to qualitatively confirm the operation of the THERMOMIR sensor, quantitative interpretation of the photothermal measurements was only possible at the device-level, as described in the following section.

\subsection{PCB integrated device characterization}
\label{subsec:mounted_device_testing}

\begin{figure}[h!]
\centering
\includegraphics[width=1.0\columnwidth]{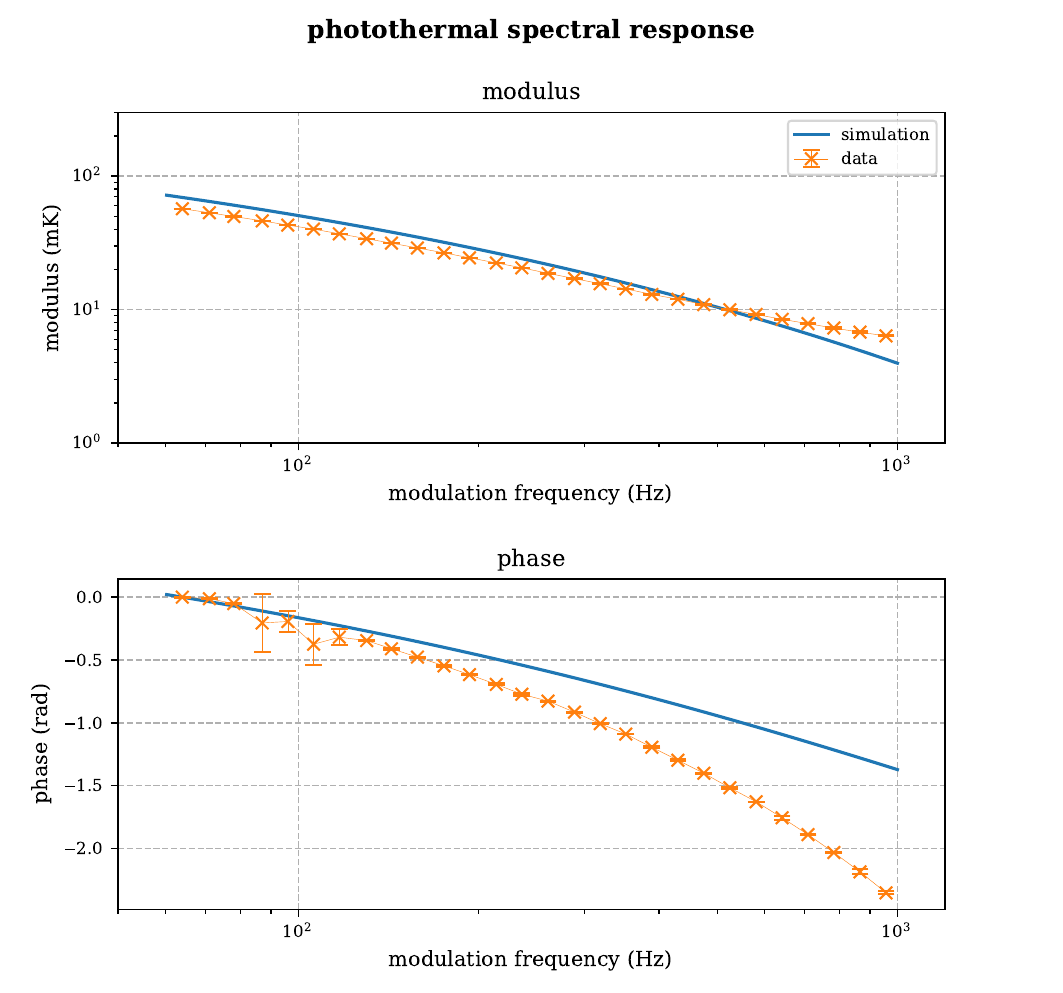}
\caption{Photothermal spectrum with respect to $f$ (optical modulation frequency) obtained using distilled water as the sample. Solid lines represent simulations and crosses represent experimentally acquired data. The top graph shows the signal amplitude, and the bottom graph shows the phase. The sample is excited at a wavelength of $9.60~\mathrm{\mu m}$ with a power of $9~\mathrm{mW}$. The data was corrected from the non-specific (or background) signal. Since the phase measurement is relative, a direct comparison with the simulation is not possible. The simulation phase is therefore offseted to match the background-corrected data (in green) at $f = 62~\mathrm{{Hz}}$. Large error-bars are due to electrical noise.}
\label{fig:photothermal_frequency_response_data_vs_simu}
\end{figure}

Once assembled on its readout PCB, the THERMOMIR sensor was fixed to a fluidic cavity connected to a manual syringe pump for liquid delivery. A tunable QCL (DAYLIGHT MIRCAT QT) was used as the optical source. It could be operated over a wavelength range from $6$ to $10~\mathrm{\mu m}$. An external mechanical chopper was used for the optical modulation. In Fig.~\ref{fig:photothermal_frequency_response_data_vs_simu} the photothermal spectrum for a distilled water sample obtained by varying the optical modulation frequency $f$ and its comparison to the simulation is presented. Excellent agreement with the simulation (solid red line) is observed at low frequencies (below $\sim 200~\mathrm{Hz}$) for the signal amplitude. It then progressively deteriorates as $f$ increases. The discrepancy at high frequencies remains unclear but it may be explained by the difference in model parameters and the actual sample properties, or by optical alignment issues.

The data in Fig.~\ref{fig:photothermal_frequency_response_data_vs_simu} were corrected from the non-specific signal, i.e., the photothermal signal detected when no sample is placed on the substrate. This contribution corresponds to the substrate intrinsic optical absorption, as well as optical leakage onto the thermometer tracks. It typically represents one tenth of the specific signal.

\begin{figure}[h!]
\centering
\includegraphics[width=1\columnwidth]{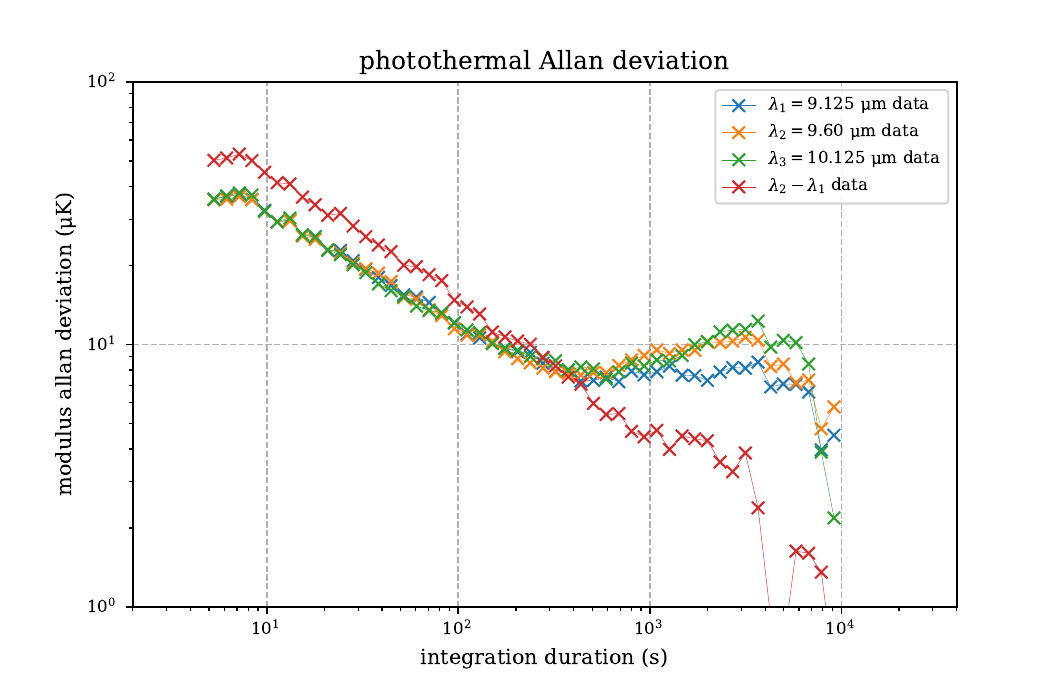}
\caption{Allan deviation of a photothermal experiment. Three $1~\mathrm{s}$ measurements were acquired at $[9.125; 9.60; 10.126]~\mathrm{\mu m}$ (blue, yellow and green) and repeated every $5~\mathrm{s}$ during $20~\mathrm{h}$. The difference between the two first measurements is also shown (red). The modulation frequency $f$ and the desposited power were kept the same ($237~\mathrm{Hz}$ and $9~\mathrm{mW}$ respectively).}
\label{fig:experimental_allan}
\end{figure}

Fig.~\ref{fig:experimental_allan} presents a time-resolved noise analysis based on Allan deviation. Three consecutive $1~\mathrm{s}$ photothermal measurements were made every $5~\mathrm{s}$ at three different wavelengths $[9.125; 9.60; 10.126]~\mathrm{\mu m}$. The modulation frequency was $237~\mathrm{Hz}$, and the deposited optical power $9~\mathrm{mW}$. The figure shows the Allan deviation for each individual measurement, as well as for the difference between the $9.60$ and $9.125~\mathrm{\mu m}$ signals. Under these operating conditions, the single-measurement noise level is approximately $\sim 30~\mathrm{\mu K}$. This value should be compared with the typical signal amplitude for a standard liquid sample~: $\sim 30~\mathrm{mK}$ (see Fig.~\ref{fig:photothermal_frequency_response_data_vs_simu}). It leads to a typical signal-to-noise ratio of approximately $1000$. This ratio can be improved or degraded by modifying the optical power level or the acquisition integration time.

The Allan curves for single measurement exhibit the expected power law behavior ($\propto 1 / \sqrt{\mathrm{\Delta t}}$ where $\Delta t~\mathrm{[s]}$ is the acquisition integration time) up to an experiment duration of approximately $\sim 100~\mathrm{s}$. For longer experiment times, the Allan deviation stabilizes or slightly increases, revealing the contribution of temporal drifts to the signal on timescales ranging from $100$ to $1000~\mathrm{s}$. Interestingly, the differential measurement suppress these drifts, enabling an increased precision for long duration experiments.

\begin{figure}[h!]
\centering
\includegraphics[width=1.0\columnwidth]{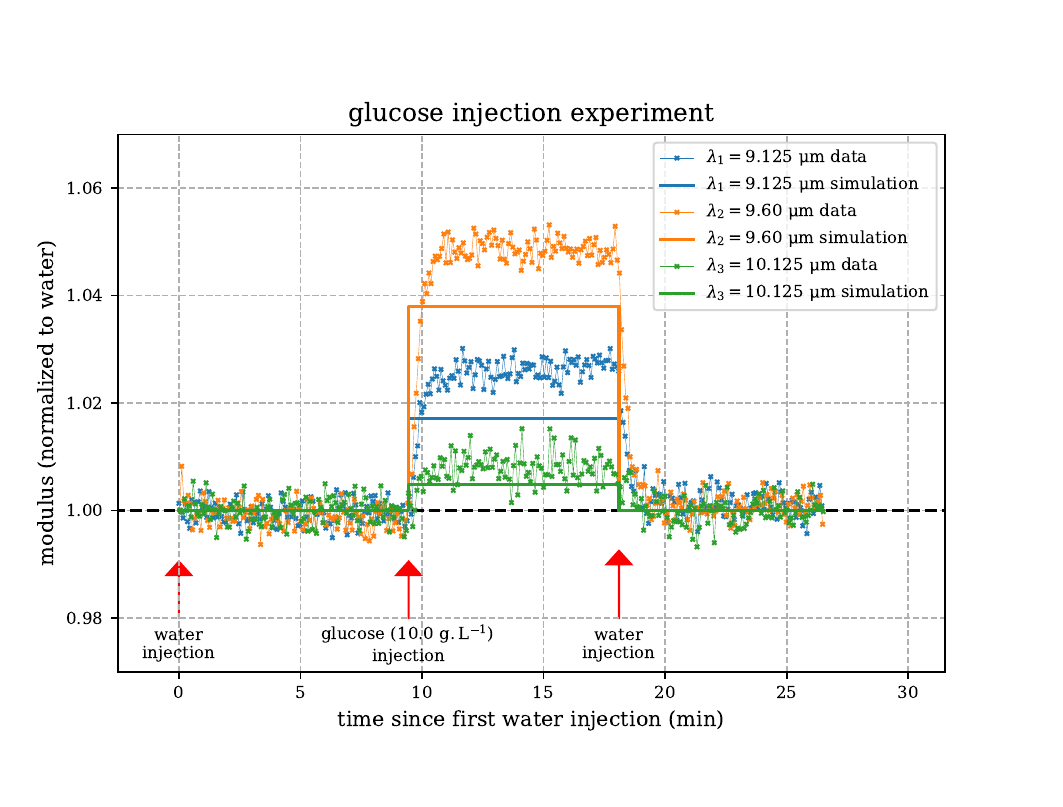}
\caption{Discrete time-spectroscopy experiment performed with THERMOMIR. The experimental conditions are similar to those of Fig.~\ref{fig:experimental_allan}. Every $5~\mathrm{s}$, three $1~\mathrm{s}$ photothermal measurements at wavelengths $[9.125; 9.60; 10.126]~\mathrm{\mu m}$ (blue, yellow and green) are acquired. The optical power and the modulation frequency $f$ remain the same ($9~\mathrm{mW}$ and $237~\mathrm{Hz}$ respectively). The experiment lasts $26~\mathrm{min}$ and starts with a water sample in the fluidic cavity. After about $9~\min$, a glucose solution ($10~\mathrm{g.L^{-1}}$ in water) is injected. Finally, after $17~\mathrm{min}$, pure water is injected back. A simulation of the expected signal (solid lines) is superimposed on the experimental data.}
\label{fig:glucose_photothermal_response}
\end{figure} 

Finally, Fig.~\ref{fig:glucose_photothermal_response} displays a typical discrete time-spectroscopy experiment performed with the integrated component. The protocol is similar as the one used for the Allan noise analysis, except that the sample is modified along the experiment. The experiment starts with distilled water, followed by the injection of a glucose solution ($10.0~\mathrm{g.L^{-1}}$ of glucose diluted in distilled water), and finally distilled water is injected again. Both experimental data, and simulations are presented. The glucose solution absorption was characterized with a standard transmission FTIR experiment (see \ref{ann:glucose_absorption}), and then used in the model described in Sec.~\ref{sec:modelling_and_design}. The photothermal data (modulus) were normalized to the signal measured during the five first minutes in distilled water.

A clear response to the sample change is observed, with stable offsets for each wavelength. Each offset reaches a specific level associated with the variation of glucose absorption between the different wavelengths. The fit with the simulation is good, with the overall trend being consistent with the experimental data. Yet, there is a systematic discrepancy (a ratio of approximately $1.3$) that is likely due to differences between the physical parameters used in the simulation and the reality (thermal and optical parameters). Eventually, the signal returns to its inital baseline when water is injected again into the fluidic chamber. It demonstrates the sensor ability to reinitialize itself.

\section{Conclusions and prospects}
\label{sec:conclusion}

A simplified detection scheme for performing IR photothermal spectroscopic measurements is reported. The device consists of a microfabricated platinum-based RTD integrated with a thin silicon substrate. The component was fabricated in a standard CMOS $200~\mathrm{mm}$ pilot line and assembled on to a readout PCB. The thin silicon substrates ($50-100~\mu m$) were transferred to the PCB by using a succession of thick holders attached to the substrate via thermo-adhesive films of different release temperatures. The components were tested at wafer and device levels. A noise analysis based on Allan deviation shows that a signal-to-noise ratio of $1000$ could be reached under standard operating conditions with a noise level around $30~\mathrm{\mu K}$. The results based on our direct photothermal measurements on distilled water sample and glucose solution indicate that our approach using the THERMOMIR device is a well-suited alternative to photoacoustic cavities, or thermal lenses, for performing IR photothermal spectroscopy. Its small form-factor, together with its ease of fabrication and implementation makes it suitable for embedded, wearable, or industrial in-line applications. Moreover, the scope of its applications can be further extended by implementing it as a sensor array, which opens up broad integration possibilities with multiple QCLs.

Several strategies can be envisioned to increase the thermometer sensitivity. For instance, the substrate thickness could be further reduced. Besides, it can also be thought as an heterogeneous structure in order to limit lateral thermal diffusion to the thermometer size. Additionally, modifying both the thermometer and beam shapes provides another degree of freedom to improve the sensitivity. For example, it would be more interesting to have line beam than a Gaussian beam shape (for a fixed optical power and $R_0$), thus, setting the thermometer tracks closer to the optical excitation. More obviously, the beam footprint $\phi_{in}$ could be reduced. While possible, it implies an increase in the complexity of the optical setup and its alignment. Furthermore, our preliminary finding suggests that the electrical track width $w$ and the spacing $s$ between the traces could be reduced down to $1~\mathrm{\mu m}$ without any significant degradation. This would increase the surface area of the thermometer within the region of interest thereby increasing the sensitivity. Finally, a material with a higher TCR than platinum, such as amorphous silicon \cite{abtew_ab_2007}, could be chosen for the RTD element.

\section*{Declaration of competing interests}

The authors declare that they have no known competing financial interests or personal relationships that could have appeared to influence the work reported in this paper.

\section*{CRediT author statement}

\textbf{Sandy Mathew~:} Conceptualization, methodology, software, investigation, writing - original draft. \textbf{Sonia Messaoudène~:} Methodology, investigation, writing - review and editing. \textbf{Adrien Poizat~:} Methodology, investigation, writing - review and editing. \textbf{\'Eléa Bourliaud~:} Investigation. \textbf{Marion Volpert~:} Methodology, investigation, writing - review and editing. \textbf{Jules Skubich~:} Methodology, investigation. \textbf{Stanislas Lhomme~:} Resources, funding acquisition. \textbf{Bertrand Bourlon~:} funding acquisition, writing - review and editing. \textbf{Kevin Jourde~:} Conceptualization, methodology, investigation, supervision, writing - original draft.

\clearpage
\newpage
\appendix

\section{Experimental setup picture}
\label{ann:characterization_benches}

\begin{figure}[h!]
\centering
\includegraphics[width=1.0\columnwidth]{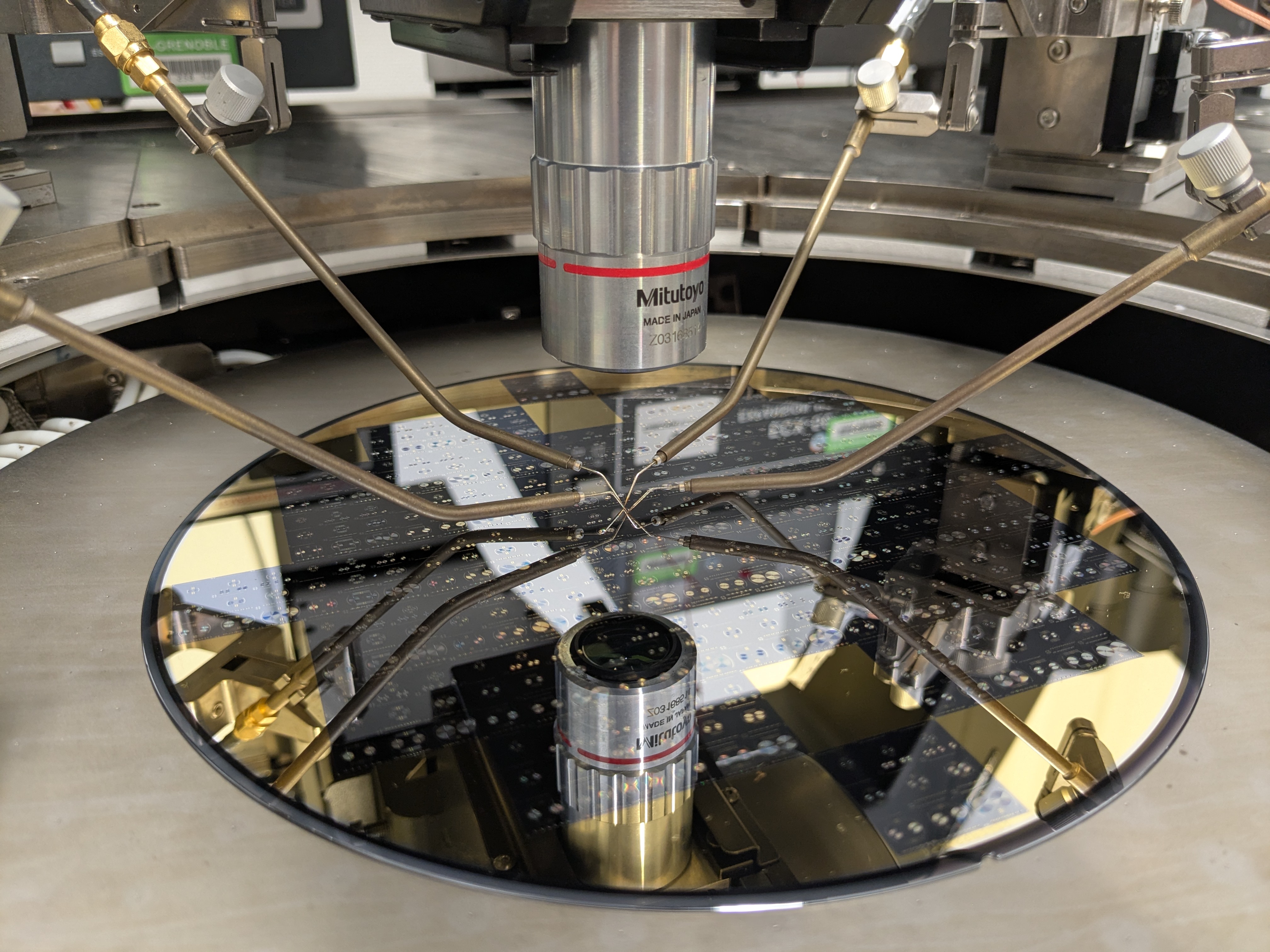}
\caption{The THERMOMIR wafer-level characterization bench for electrical testing. The automated prober is visible in the background with a wafer loaded on the chuck. The wafer is probed both electrically via four electrical probes set all around the chuck.}
\label{fig:wafer_level_bench}
\end{figure} 

\begin{figure}[h!]
\centering
\includegraphics[width=1.0\columnwidth]{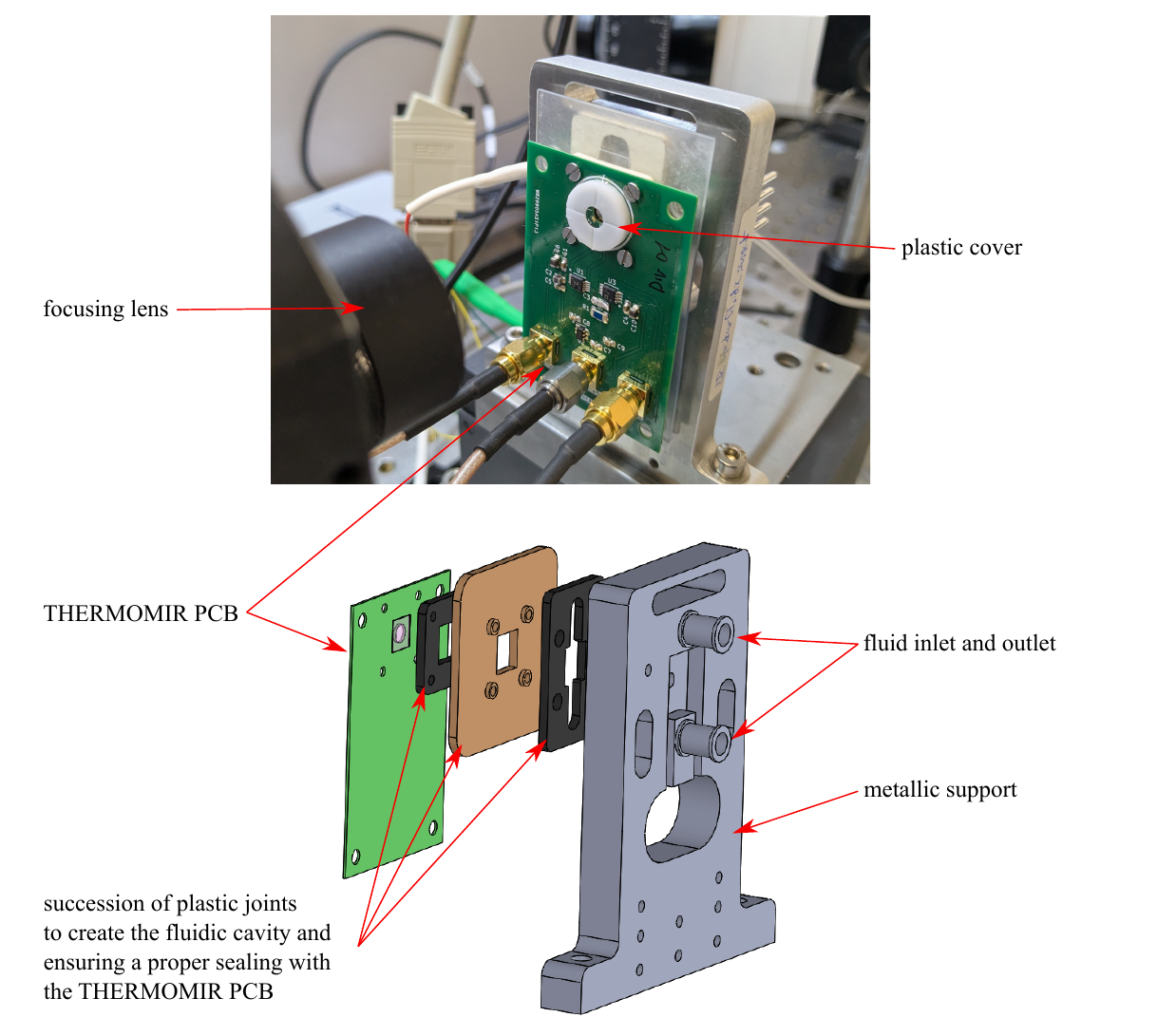}
\caption{The THERMOMIR device-level characterization setup. On the top is a picture showing the THERMOMIR readout PCB front face, while mounted on the fluidic cavity. On the bottom is an exploded vision of the 3D assembly. It shows the same elements as in the picture but from the opposite direction. The two images are annotated to distinguish the different parts.}
\label{fig:device_level_bench}
\end{figure} 

\clearpage
\newpage
\section{MIR glucose absorption in water}
\label{ann:glucose_absorption}

\begin{figure}[h!]
\centering
\includegraphics[width=1.0\columnwidth]{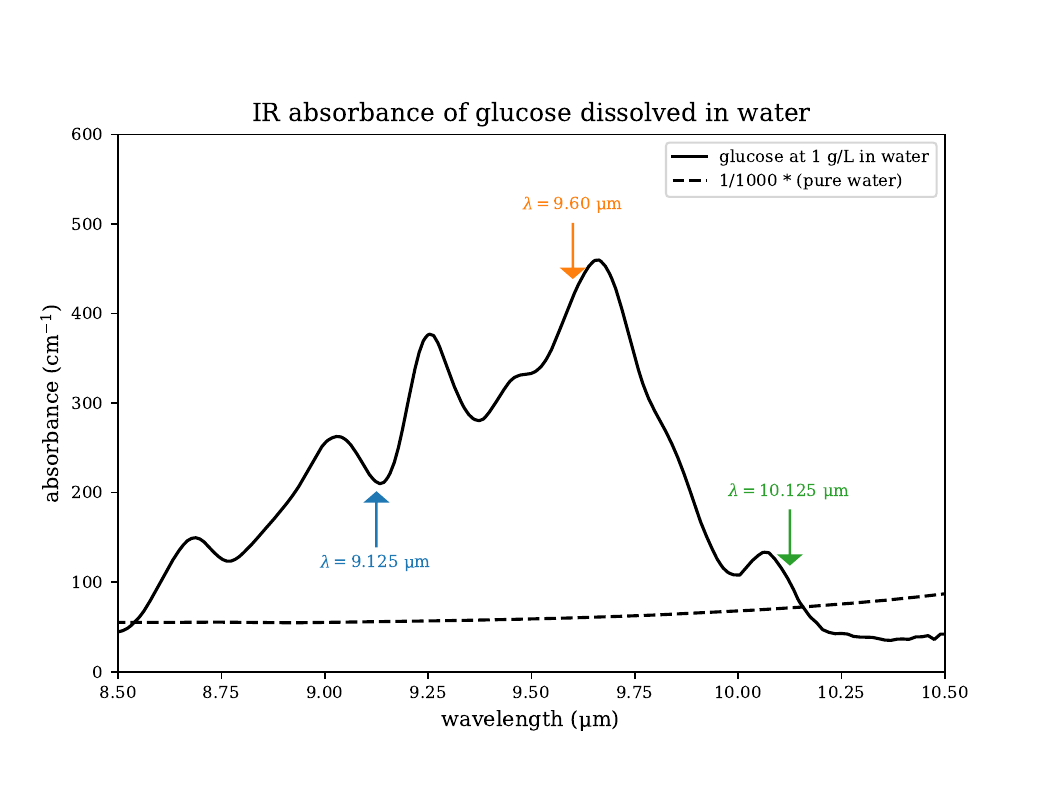}
\caption{This graph shows the IR absorbance  (between $8.5$ and $10.5~\mathrm{\mu m}$) of glucose dissolved in distilled water and distilled water alone (scaled by a factor of $1/1000$). The arrows indicate the wavelengths used in the characterization experiments. The data was obtained with a conventional transmission FTIR instrument.}
\label{fig:glucose_absorption}
\end{figure} 

\bibliographystyle{elsarticle-num}
\bibliography{thermomir_paper}

\end{document}